\documentclass[12pt]{article}
\usepackage{amssymb,amsmath,epsfig}

\begin{document}

\title{\bf Collapse and Expansion of Scalar Thin-Shell for a Class of Black Holes}
\author{M. Sharif \thanks {msharif.math@pu.edu.pk} and Faisal Javed
\thanks{faisalrandawa@hotmail.com}\\
Department of Mathematics, University of the Punjab,\\
Quaid-e-Azam Campus, Lahore-54590, Pakistan.}

\date{}
\maketitle

\begin{abstract}
This paper investigates the dynamics of thin-shell in the presence
of perfect fluid as well as scalar field. We formulate the equations
of motion using Israel thin-shell formalism by taking the interior
and exterior regions of Schwarzschild, Kerr as well as Kerr-Newmann
black hole. We find numerical solutions of equations of motion and
effective potential to analyze the scalar shell for collapse and
expansion. It is found that the rate of collapse and expansion of
scalar shell through shell's radius depends on charge and rotation
parameters. We conclude that the massive scalar shell leads to
collapse of thin-shell, while massless scalar shell indicates both
collapse as well as expansion.
\end{abstract}

\section{Introduction}

Geon is a notion of electromagnetic-gravitational field that is held
together in a confined region due its own gravity. This entity was
introduced by Wheeler et al. \cite{1} who established particle-like
solutions from classical electromagnetic field coupled to general
relativity. They used the resulting solutions to analyze the scalar
field. Bergmann and Leipnik \cite{2} studied solutions of the field
equations in the presence of scalar field for Schwarzschild black
hole (BH). In general relativity, scalar field appears in the low
energy limit of string theory \cite{3}. There is no experimental
evidence for the existence of such particles that are associated
with the scalar field. Harada et al. \cite{4} studied that
gravitational collapse of compact objects in scalar-tensor theories
predict scalar field as a source of gravitational waves that can be
identified by the advanced detectors.

The study of cosmological as well as astrophysical objects composed
of scalar field has been the subject of great interest for many
researchers. Kaup \cite{5} was the pioneer to investigate the
geometrical configuration for complex massive scalar field. Ruffini
and Bonazzola \cite{6} examined spherical geometries and found the
equilibrium conditions for boson stars solutions. Seidel and Suen
\cite{7} studied the dynamical evolution of compact stars associated
with scalar field and found that boson stars have stable
configurations for several values of the scalar fields. These
results provide information corresponding to existence as well as
formation of boson stars. Jetzer \cite{8} investigated the
equilibrium configuration of boson stars and discussed their
dynamical instability. Many people \cite{9} investigated the nature
of spacetime singularity for massless scalar field with spherical
geometry. Siebel et al. \cite{10} analyzed self-gravitating objects
composed of massless scalar field and found that boson stars either
oscillate or undergo collapse to form a BH. Bhattacharya et al.
\cite{11} studied the collapse of spherical stars associated with
massless scalar field and examined appropriate condition for the
existence of a class of non-singular models.

The smooth matching of interior and exterior spacetimes helps to
evaluate exact solutions of the field equations. This is also a key
aspect to study the boundary of BHs, gravitational waves and
contribution of matter at thin-shell. Israel \cite{12} developed
thin-shell formalism to investigate the dynamics of fluid
configuration at thin-shell. It is observed that the presence of
thin layer of matter at thin-shell leads to jump discontinuity
across the boundary of interior and exterior regions. In order to
analyze thin-shell, Israel formalism has widely been used \cite{13}.

De La Cruz and Israel \cite{14} generalized Israel thin-shell
formalism for charged thin-shell in the absence of pressure. Kuchar
\cite{15} examined the charged thin-shell problem by considering
polytropic equation of state (EoS). Chase \cite{16} investigated the
instability of spherically symmetric charged fluid shell. Boulware
\cite{17} analyzed charged thin-shell and found that collapse
produces a naked singularity when matter (with negative density) is
located at thin-shell. N$\acute{u}\tilde{n}$ez \cite{18} studied
spherically symmetric massive shell and found that the shell
oscillates radially around a central compact object.
N$\acute{u}\tilde{n}$ez et al. \cite{19} analyzed the effect of
scalar field for Schwarzschild BH and found that massive scalar
field leads to collapse of the shell when both pressure and
gravitational mass are proportional. Pereira and Wang \cite{20} used
thin-shell formalism to study the non-rotating shell constructed
from two cylindrical regions and observed the behavior of collapsing
shell. Sharif and his collaborators \cite{21} used this formalism to
investigate spherical as well as planar collapse. Sharif and Abbas
\cite{22} explored the dynamics of charged scalar thin-shell and
concluded that for both (massless and massive scalar fields) shell
can expand to infinity or collapse to zero size forming a curvature
singularity. Sharif and Iftikhar \cite{23} explored a class of
regular BHs and found that massless scalar shell leads to expansion,
collapse and equilibrium structure while the massive case leads to
collapse only.

This work is devoted to study the effect of scalar field on the
dynamics of thin-shell using Israel formalism for a class of BHs.
The paper is organized as follows. In section \textbf{2}, we
construct the equations of motion by applying Israel thin-shell
formalism for Schwarzschild, Kerr and Kerr-Newmann BHs. Section
\textbf{3} explores massless and massive scalar fields to
investigate these equations of motion. Finally, we summarize our
results in the last section.

\section{Equations of Motion}

In this section, we construct thin-shell for Schwarzschild, Kerr and
Kerr-Newmann BHs. We consider a hypersurface ($\Sigma$) that divides
a spherically symmetric spacetime into two four-dimensional
manifolds $N^{+}$ and $N^{-}$ representing interior and exterior
regions, respectively. The line element for interior and exterior
regions of the Schwarzschild, Kerr and Kerr-Newmann BHs is given as
\begin{eqnarray}\nonumber
ds^2_\pm&=&-F_{\pm}(R)dt^2-\left(\frac{4m_{\pm}R}
{A}-\frac{2Q^2_{\pm}}{A}\right)a\sin^2\theta
dtd\phi+\left(\frac{A}{B}\right)dr^2+A d\theta^2\\\label{1}&+&
\left(R^2+a^2+\frac{2m_{\pm}a^2R\sin^2\theta}{A}-
\frac{a^2Q^2_{\pm}\sin^2\theta}{A}\right)\sin^2\theta d\phi^2,
\end{eqnarray}
where
\begin{eqnarray}\nonumber
A=R^2+a^2cos^2\theta, \quad B=R^2-2m_{\pm}R+a^2+Q^2_{\pm},
\end{eqnarray}
here $m_{\pm}$, $Q_{\pm}$ and $a$ denote the mass, charge and
rotation parameters, respectively. Moreover, we assume that interior
region contains more mass than the exterior region, i.e., $m_{+}\neq
m_-$, while the charge is uniformly distributed in both regions,
i.e., $Q_+=Q_-=Q$. The explicit forms of the defining parameters
corresponding to different BHs are given in
Table \textbf{1}.\\\\
\textbf{Table 1:} A Class of BHs.
\begin{table}[htbp]
\begin{tabular}{|l||c|} \hline
\textbf{Name of BH} & $F(R)$  \\ \hline \textbf{Schwarzschild} &
$F_{\pm1}(R)=\left(1-\frac{2m_{\pm}}{R}\right)$, $a=0$, $Q_{\pm}=0$
\\\hline \textbf{Kerr} &
$F_{\pm2}(R)=\left(1-\frac{2m_{\pm}R}{A}\right)$, $Q_{\pm}=0$
\\\hline \textbf{Kerr-Newmann} &
$F_{\pm3}(R)=\left(1-\frac{2m_{\pm}R}{A}+\frac{Q^2_{\pm}}{A}\right)$
\\\hline
\end{tabular}
\end{table}

We apply the intrinsic coordinates $\xi^{i}=(\tau,\theta,\phi)$ over
$\Sigma$ at $R=R(\tau)$. Consequently, Eq.(\ref{1}) yields
\begin{eqnarray}\nonumber
ds^2_\pm&=&\left[-F_{\pm}(R)+\frac{A}{B}\left(\frac{dR}{d\tau}
\right)^2\left(\frac{d\tau}{dt}\right)^2\right]dt^2-\left(\frac{4m_{\pm}R}
{A}-\frac{2Q^2_{\pm}}{A}\right)a\sin^2\theta d\phi
dt\\\label{2aa}&+&A d\theta^2+
\left(R^2+a^2+\frac{2m_{\pm}a^2R\sin^2\theta}{A}-
\frac{a^2Q^2_{\pm}\sin^2\theta}{A}\right)\sin^2\theta d\phi^2,
\end{eqnarray}
where $\tau$ is the proper time. The corresponding induced metric
for hypersurface is defined as
\begin{equation}\label{3aa}
ds^2=-d\tau^2+R^2(\tau)\left(d\theta^2+\sin^2\theta d\phi^2\right).
\end{equation}
Comparing Eqs.(\ref{2aa}) and (\ref{3aa}), we obtain
\begin{eqnarray}\nonumber
\left[F_{\pm}(R)-\frac{A}{B}\left(\frac{dR}{d\tau}
\right)^2\left(\frac{d\tau}{dt}\right)^2\right]^\frac{1}{2}dt=(d\tau)_{\Sigma}.
\end{eqnarray}
The outward unit normals $n_{\alpha}^{\pm}$ corresponding to
interior as well as exterior region are
$n_{\alpha}^{\pm}=(n_{0},n_{1},0,0)$, where
\begin{eqnarray}\nonumber
n_{0}&=&-\dot{R}\left(\frac{B + A \dot{R}^2}{B
F_{\pm}}\right)^{-\frac{1}{2}}\left[\frac{B}{A}+\left\{A B \dot{R}^2
F_{\pm} \left(-A \left(a^2 + R^2\right) +
a^2\sin^2\theta\right.\right.\right.
\\\nonumber &\times&\left.\left.\left.\left(Q^2 - 2 m_{\pm} R\right)
\right)\right\}\left\{\left(B + A \dot{R}^2\right) \left(A^2 F_{\pm}
\left(a^2 + R^2\right) - a^2 \left(Q^2 - 2 m_{\pm}
R\right)\right.\right. \right.\\\nonumber&\times&
\left.\left.\left.\left(A F_{\pm} - 4 Q^2 + 8 m_{\pm} R\right)
\sin^2\theta\right)\right\}^{-1}\right]^{-\frac{1}{2}},
\\\nonumber
n_{1}&=&\left[\frac{B}{A}+\left\{A B \dot{R}^2 F_{\pm} \left(-A
\left(a^2 + R^2\right) + a^2\sin^2\theta\left(Q^2 - 2 m_{\pm}
R\right)
\right)\right\}\right.\\\nonumber&\times&\left.\left\{\left(B + A
\dot{R}^2\right) \left(A^2 F_{\pm} \left(a^2 + R^2\right) - a^2
\left(Q^2 - 2 m_{\pm} R\right)\right.\right.
\right.\\\nonumber&\times& \left.\left.\left.\left(A F_{\pm} - 4 Q^2
+ 8 m_{\pm} R\right)
\sin^2\theta\right)\right\}^{-1}\right]^{-\frac{1}{2}}.
\end{eqnarray}
Here, dot denotes derivative with respect to $\tau$. The extrinsic
curvatures joining two sides of the shell are defined as
\begin{equation}\label{4}
K_{ij}^{\pm}=-n_{\beta}^{\pm}\left(\frac{d^2x_{\pm}^\beta}{d\xi^id\xi^j}
+\Gamma^\beta_{\mu\nu}\frac{dx^{\mu}_{\pm} dx^{\nu}_{\pm}} {d\xi^{i}
d\xi^{j}}\right),\quad \mu,\nu,\beta=0,1,2,3.
\end{equation}

The discontinuity of extrinsic curvature appears due to the
existence of thin layer of matter on $\Sigma$. The dynamics of this
thin-shell is observed by using the field equations at $\Sigma$,
i.e., by the Lanczos equations
\begin{equation}\label{5}
S_{ij}=\frac{1}{8\pi}\{g_{ij}K-[K_{ij}]\},\quad i,j=0,2,3,
\end{equation}
where $[K_{ij}]=K^{+}_{ij}-K^{-}_{ij}$ and
$K=tr[K_{ij}]=[K^{i}_{i}]$. The Lanczos equations for spherical
thin-shell reduces to
\begin{eqnarray}\label{6}
\sigma=\frac{-1}{4\pi}[K_{\theta}^{\theta}], \quad
p=\frac{1}{8\pi}\left\{
[K_{\tau}^{\tau}]+[K_{\theta}^{\theta}]\right\}.
\end{eqnarray}
In the absence of surface energy density ($\sigma$) and pressure
($p$), the connection between these geometries is referred as a
boundary surface, otherwise it is known as thin-shell. The
corresponding surface energy density can be expressed as
\begin{eqnarray}\nonumber
\sigma &=&\frac{B}{8\pi r A}\left[\frac{B}{A}+\left\{A B \dot{R}^2
F_{\pm} \left(-A \left(a^2 + R^2\right) + a^2\sin^2\theta\left(Q^2 -
2 m_{\pm} R\right) \right)\right\}\right.\\\nonumber &\times&\left.
\left\{\left(B + A \dot{R}^2\right) \left(A^2 F_{\pm} \left(a^2 +
R^2\right) - a^2 \left(Q^2 - 2 m_{\pm} R\right)\right.\right.
\right.\\\label{7}&\times& \left.\left.\left.\left(A F_{\pm}- 4 Q^2
+ 8 m_{\pm}
R\right)\sin^2\theta\right)\right\}^{-1}\right]^{-\frac{1}{2}}.
\end{eqnarray}
The above equation can be rewritten as
\begin{eqnarray}\label{8}
\dot{R}^2 + V_{eff}(R)=0,
\end{eqnarray}
where
\begin{eqnarray}\nonumber
V_{eff}(R)&=&\left\{B \left(B - 64 A \pi^2 r^2 \sigma^2\right)
\left(-A^2 F_{\pm} \left(a^2 + r^2\right) + a^2 \left(Q^2 - 2
m_{\pm} r\right) \right.\right.\\\nonumber&\times&
\left.\left.\left(A F_{\pm} - 4 Q^2 + 8 m_{\pm} r\right)
\sin^2\theta\right)\right\}\left\{A^3 B F_{\pm} \left(a^2 +
r^2\right) - a^2 A \right.\\\nonumber&\times&\left.\left(Q^2 - 2
m_{\pm} r\right) \left(B \left(A F_{\pm} - 4 Q^2 + 8 m_{\pm}
r\right) + 256 A \pi^2 r^2
\right.\right.\\\label{9}&\times&\left.\left.\left(Q^2 - 2 m_{\pm}
r\right) \sigma^2\right) \sin^2\theta\right\}^{-1}.
\end{eqnarray}
It is observed that Eq.(\ref{8}) satisfies the energy conservation
law, i.e., the sum of the component of kinetic $(\dot{R}^2)$ and
potential $(V_{eff}(R))$ energies vanishes at any time. The
effective potential for Schwarzschild, Kerr and Kerr-Newmann BHs
turns out to be
\begin{eqnarray}\label{10}
V_{eff1}(R) &=&-F_{\pm1} + 16\pi^2 R^2 \sigma^2,
\\\nonumber
V_{eff2}(R)&=&\left\{B \left(B - 16 A \pi^2 R^2 \sigma^2\right)
\left(A^2 F_{\pm2} \left(a^2 + R^2\right) + 2 a^2 m_{\pm} R \left(A
F_{\pm2} \right.\right.\right.\\\nonumber&+&\left.\left.\left. 8
m_{\pm} R\right) \sin^2\theta\right)\right\}\left\{A^3 B F_{\pm2}
\left(a^2 + R^2\right) + 2 a^2 A m_{\pm} R \left(A B F_{\pm2}
\right.\right.\\\label{11}&+&\left.\left.8 B m_{\pm} R - 128 A
m_{\pm} \pi^2 R^3 \sigma^2\right) \sin^2\theta\right\}^{-1},
\\\nonumber
V_{eff3}(R)&=&\{(B (B - 64 A \pi^2 R^2 \sigma^2) (-A^2 F_{\pm3} (a^2
+ R^2) + a^2(Q^2 - 2 m_{\pm} R)
\\\nonumber&\times&(A F_{\pm3}-4 Q^2 +
8 m_{\pm} R) \sin^2\theta))\} \{(A^3 B F_{\pm3} (a^2 + R^2) - a^2 A
\\\nonumber&\times&(Q^2 - 2 m_{\pm} R)(B (A F_{\pm3} - 4 Q^2 + 8
m_{\pm} R) + 256 A \pi^2 R^2\\\label{12}&\times&(Q^2 - 2 m_{\pm} R)
\sigma^2) \sin^2\theta\}^{-1}.
\end{eqnarray}

\section{Analysis of Equations of Motion}

Here we investigate the scalar shell and its dynamical behavior for
a class of BHs. We evaluate velocity of the scalar shell and its
effective potential with respect to stationary observer. Also, we
discuss the effect of charge as well as rotation parameter on the
motion of scalar shell. For this purpose, we formulate a relation
between surface energy density and the mass of thin-shell. The
surface energy density and pressure follow the conservation equation
\begin{equation}\label{13}
p \frac{d\Delta}{d\tau}+\frac{d}{d\tau}(\sigma \Delta)=0,
\end{equation}
where $\Delta=4\pi R^2$ represents area of the shell. Consequently,
the conservation equation for $p=p(\sigma,R)$ takes the form
\begin{equation}\label{14}
\sigma'+\frac{2}{R}\left[\sigma+p(\sigma,R)\right]=0.
\end{equation}
This equation can be solved by using EoS, $p=k\sigma$, yielding
\begin{eqnarray}\label{15}
\sigma=\sigma_{0}\left(\frac{R_{0}}{R}\right)^{2(k+1)},
\end{eqnarray}
where $k$ is a constant and $R_{0}$ represents initial position of
the shell at $\tau=\tau_{0}$ while $\sigma_0$ denotes surface
density of the shell at $R_{0}$. Using the above equation, mass of
the shell becomes
\begin{eqnarray}\label{16}
M=4\pi\sigma_{0}\left(\frac{R_{0}^{2(k+1)}}{R^{2k}}\right).
\end{eqnarray}

In the following, we briefly discuss the dynamics of thin-shell
using equations of motion.

\subsection{The Schwarzschild Black Hole}

The effective potential for Schwarzschild BH is given by
\begin{eqnarray}\label{17}
V_{eff1}(R) = -1+ \frac{M^2}{R^2} +\frac{2m_{\pm}}{R}.
\end{eqnarray}
The corresponding equations of motion of the shell becomes
\begin{eqnarray}\label{18}
\dot{R} =\pm \left[1- \frac{M^2}{R^2}
-\frac{2m_{\pm}}{R}\right]^\frac{1}{2},
\end{eqnarray}
here $\pm$ represents expansion and collapse of thin-shell. In
Figure \textbf{1}, the left and right graphs describe the shell
velocity for $\dot{R}>0$ and $\dot{R}<0$, respectively. The
expansion (collapse) of the scalar shell is shown in left (right)
graph, while blue and red curves correspond to interior and exterior
regions. These graphs indicate that thin-shell velocity decreases
positively (left plot) and increases negatively (right plot). Also,
it is found that the velocity of interior region is greater than the
exterior.
\begin{figure}\centering
\epsfig{file=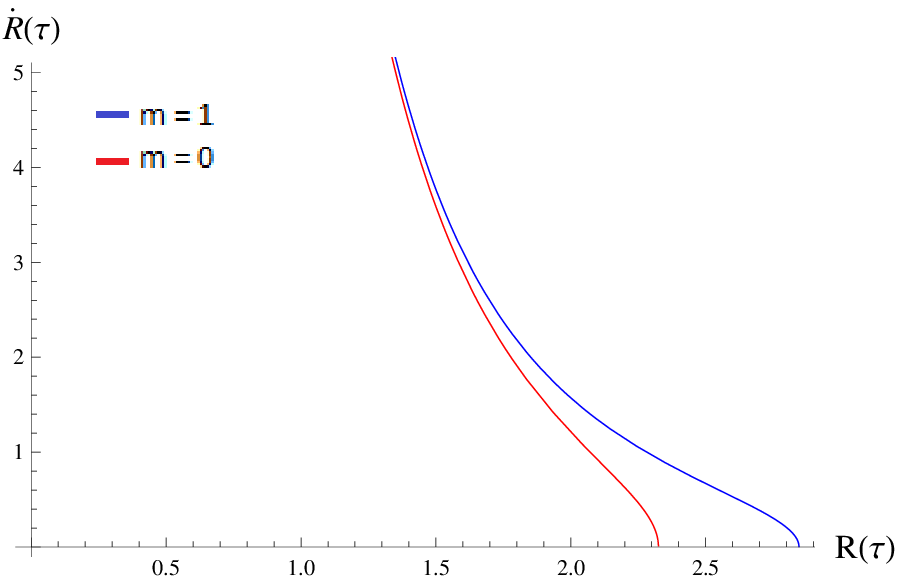,width=.5\linewidth}\epsfig{file=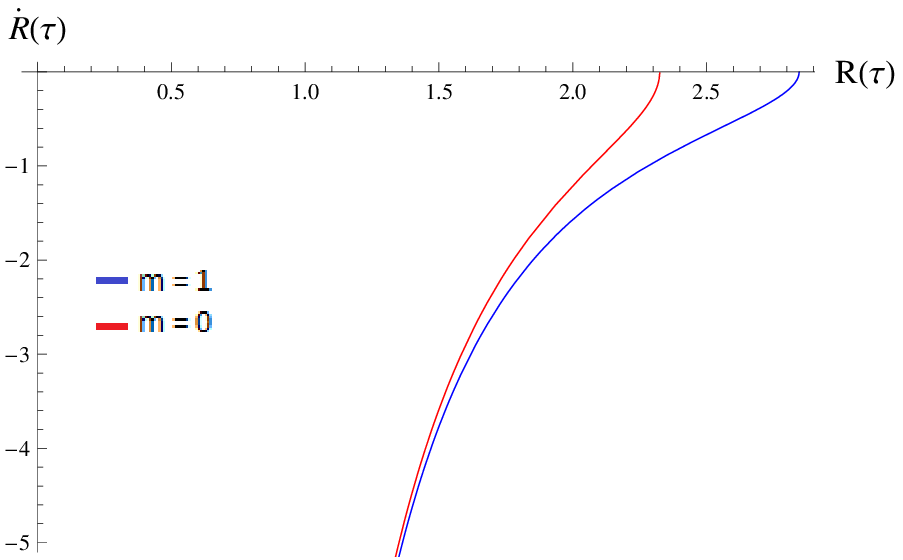,width=.5\linewidth}
\caption{Behavior of thin-shell velocity corresponding to $R$ for
Schwarzschild BH with $R_0=\sigma_0=k=1$.}
\end{figure}

\subsection{The Kerr Black Hole}

For Kerr BH, the effective potential takes the following form
\begin{eqnarray}\nonumber
V_{eff2}(R)&=&\left\{B \left(B-\frac{A M^2}{R^2}\right)
\left(A^2F_{\pm} \left(a^2 + R^2\right)+2 a^2 m_{\pm} R\sin^2\theta
\right.\right.\\\nonumber&\times&\left.\left. \left(A F_{\pm} +8
m_{\pm} R\right) \right)\right\}\left\{ \left(A^3 B F_{\pm}\left(a^2
+R^2 \right)+2AR a^2 m_{\pm}
\right.\right.\\\label{19}&\times&\left.\left(A B F_{\pm}-\frac{8 A
m_{\pm} M^2}{R} +8 B m R\right)\sin^2\theta\right\}^{-1}.
\end{eqnarray}
In this case, the respective equations of motion becomes
\begin{eqnarray}\nonumber
\dot{R}&=&\pm\left[-\left\{B \left(B-\frac{A M^2}{R^2}\right)
\left(A^2F_{\pm} \left(a^2 + R^2\right)+2 a^2 m_{\pm} R\sin^2\theta
\right.\right.\right.\\\nonumber&\times&\left.\left.\left. \left(A
F_{\pm} +8 m_{\pm} R\right) \right)\right\}\left\{ \left(A^3 B
F_{\pm}\left(a^2 +R^2 \right)+2AR a^2 m_{\pm}
\right.\right.\right.\\\label{20}&\times&\left.\left.\left(A B
F_{\pm}-\frac{8 A m_{\pm} M^2}{R} +8 B m
R\right)\sin^2\theta\right\}^{-1}\right]^{\frac{1}{2}}.
\end{eqnarray}
The plots for shell's velocity is shown in Figure \textbf{2}. It is
observed that the behavior of velocity in the presence of rotation
parameter remains the same as for the Schwarzschild BH.
\begin{figure}\centering\label{2}
\epsfig{file=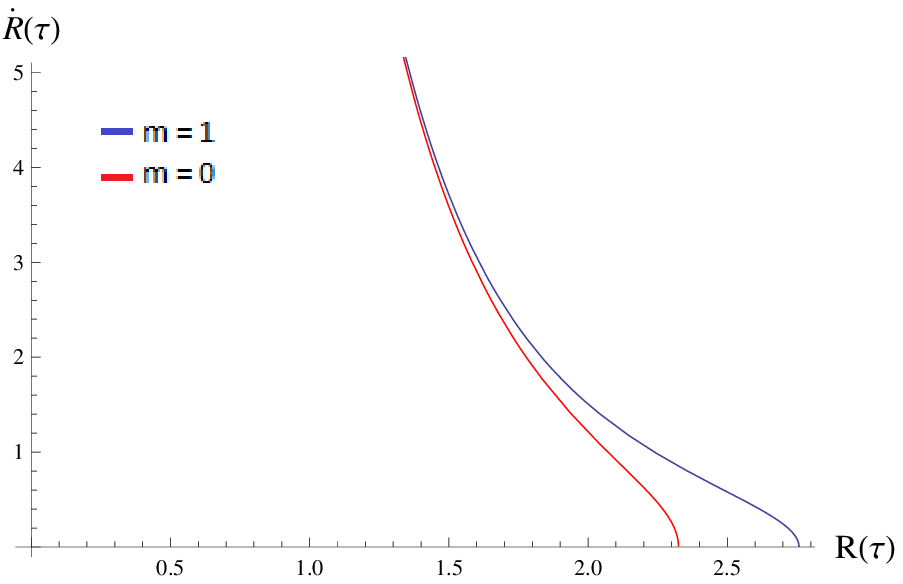,width=.5\linewidth}\epsfig{file=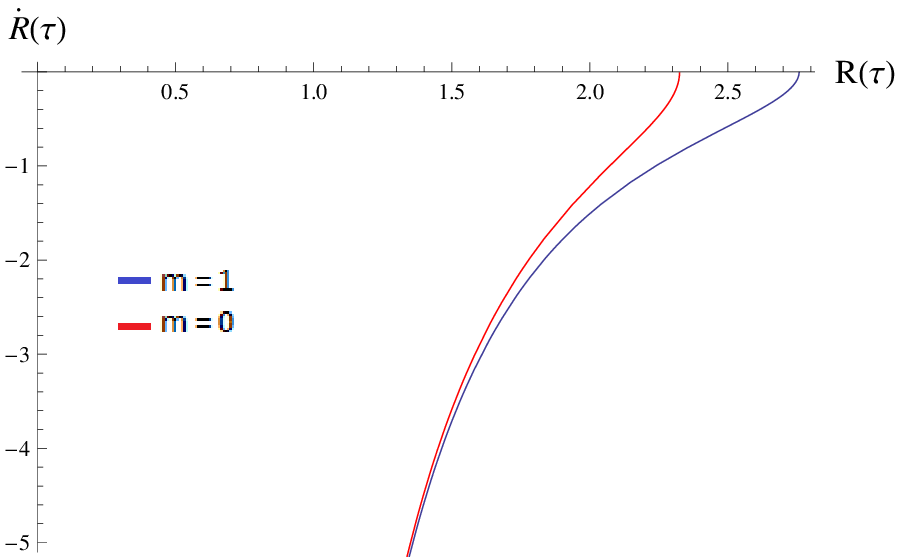,width=.5\linewidth}
\caption{Behavior of thin-shell velocity corresponding to $R$ for
Kerr BH with $R_0=\sigma_0=k=a=1$.}
\end{figure}

\subsection{The Kerr-Newmann Black Hole}

This BH is a generalization of the Kerr BH. The corresponding
effective potential is
\begin{eqnarray}\nonumber
V_{eff3}(R)&=&-\left\{B \left(B-\frac{4A M^2}{R^2}\right)
\left(-A^2F_{\pm} \left(a^2 + R^2\right)+a^2\left(Q^2-2
m_{\pm}R\right) \right.\right.\\\nonumber&\times&\left.\left.
\left(A F_{\pm}-4Q^2 +8 m_{\pm} R\right)\sin^2\theta
\right)\right\}\left\{ \left(A^3 B F_{\pm}\left(a^2 +R^2
\right)-a^2A \right.\right.\\\nonumber&\times&\left.\left(\frac{16 A
\left(Q^2-2m_{\pm}R\right) M^2}{R^2} + B\left(A F_{\pm}-4Q^2+8
m_{\pm} R\right)\right)\right.\\\label{21}&
\times&\left.\left(Q^2-2m_{\pm}R\right)\sin^2\theta \right\}^{-1}.
\end{eqnarray}
The equations of motion for this BH takes the form
\begin{figure}\centering\label{3}
\epsfig{file=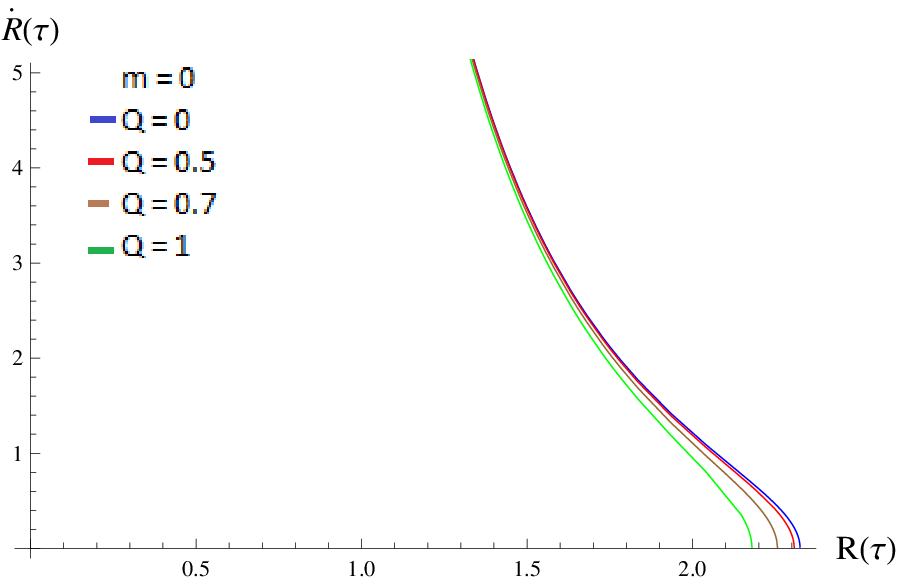,width=.5\linewidth}\epsfig{file=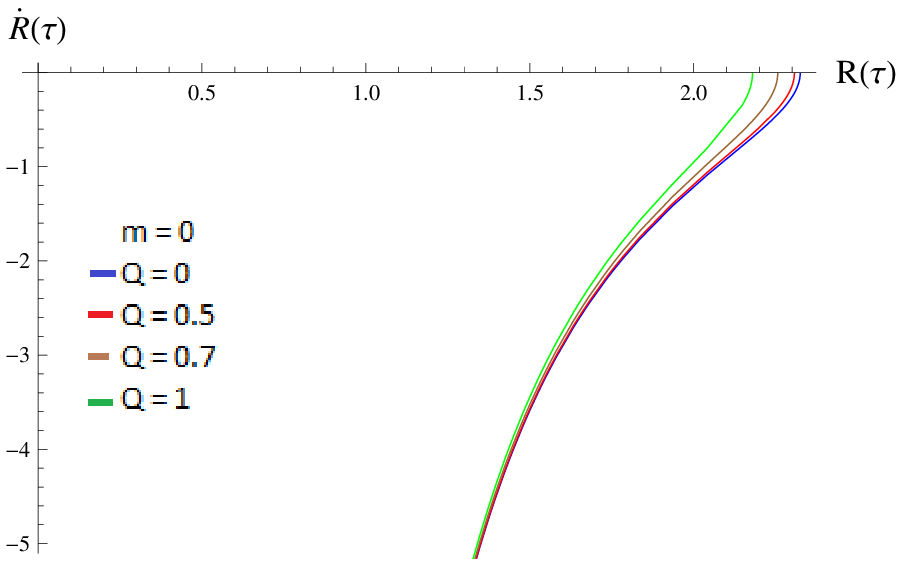,width=.5\linewidth}
\epsfig{file=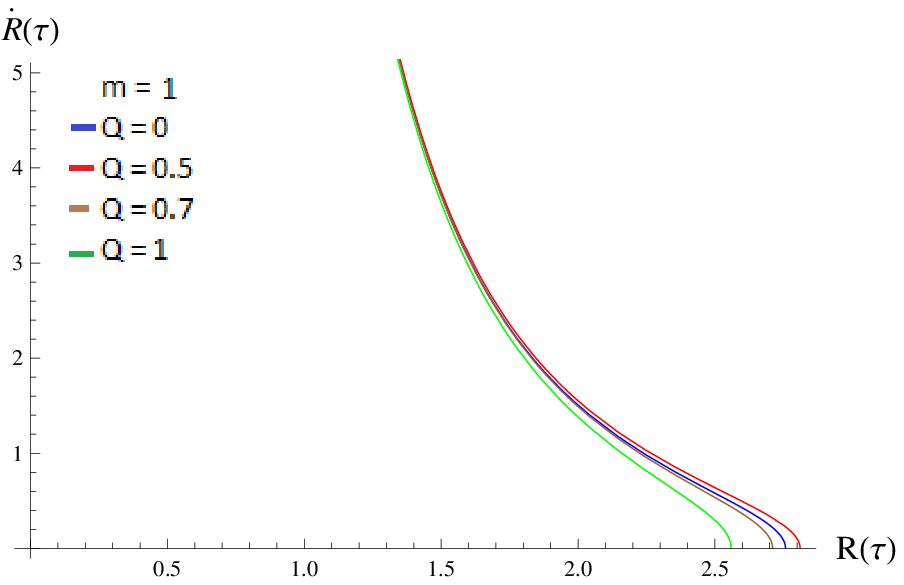,width=.5\linewidth}\epsfig{file=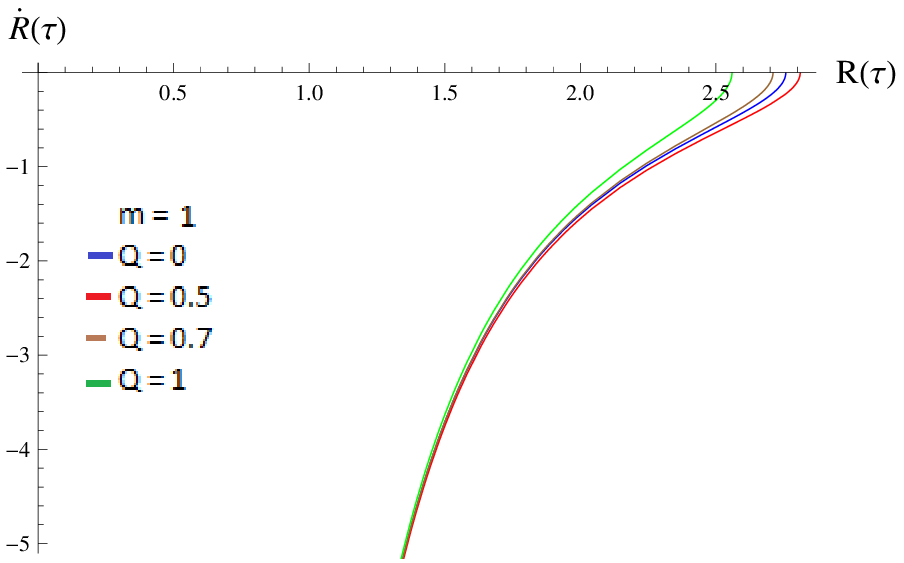,width=.5\linewidth}
\caption{Behavior of thin-shell velocity corresponding to $R$ for
Kerr-Newmann BH with $R_0=\sigma_0=k=a=1$.}
\end{figure}
\begin{eqnarray}\nonumber
\dot{R}&=&\pm\left[\left\{B \left(B-\frac{4A M^2}{R^2}\right)
\left(-A^2F_{\pm} \left(a^2 + R^2\right)+a^2\left(Q^2-2
m_{\pm}R\right)
\right.\right.\right.\\\nonumber&\times&\left.\left.\left. \left(A
F_{\pm}-4Q^2 +8 m_{\pm} R\right)\sin^2\theta \right)\right\}\left\{
\left(A^3 B F_{\pm}\left(a^2 +R^2 \right)-a^2A
\right.\right.\right.\\\nonumber&\times&\left.\left.\left(\frac{16 A
\left(Q^2-2m_{\pm}R\right) M^2}{R^2} + B\left(A F_{\pm}-4Q^2+8
m_{\pm} R\right)\right)\right.\right.\\\label{22}&
\times&\left.\left.\left(Q^2-2m_{\pm}R\right)\sin^2\theta
\right\}^{-1}\right]^{\frac{1}{2}}.
\end{eqnarray}
Figure \textbf{3} (upper panel is for exterior while lower is for
interior region) indicates that the velocity of thin-shell depends
on charge but there is no contribution of rotation parameter. When
we increase charge, the velocity decreases positively for both
spacetimes as shown in the left plot of upper as well as lower
panel. It is observed that velocity increases negatively for these
geometries as shown in the right plot of upper as well as lower
panel.

\subsection{Dynamics of Scalar Shell}

Here we examine the dynamical behavior of the scalar shell. For this
purpose, we use a transformation
$\left(u_{a}=\frac{\psi_{,a}}{\sqrt{\psi_{,b}\psi^{,b}}}\right)$
\cite{19}, which relates surface energy density and pressure of a
perfect fluid with potential function $V(\psi)$ and derivative of
the scalar field. The corresponding surface energy density and
pressure are obtained as follows
\begin{eqnarray}\label{23}
\sigma=-\frac{1}{2}\left[\psi_{,b}\psi^{,b}-2V(\psi)\right],\quad
p=-\frac{1}{2}\left[\psi_{,b}\psi^{,b}+2V(\psi)\right],
\end{eqnarray}
where $V(\psi)=M^2\psi^2$. The stress-energy tensor in terms of
scalar field is defined as
\begin{eqnarray}\nonumber
S_{ij}=\nabla_{i}\psi
\nabla_{j}\psi-\eta_{ij}\left[\frac{1}{2}(\nabla
\psi)^2-V(\psi)\right].
\end{eqnarray}
Since, the induced metric is a function of proper time $\tau$, so
$\psi$ depends on $\tau$. Consequently, Eq.(23) yields
\begin{eqnarray}\label{24}
\sigma=\frac{1}{2}\left[\dot{\psi}^2+2V(\psi)\right], \quad
p=\frac{1}{2}\left[\dot{\psi}^2-2V(\psi)\right].
\end{eqnarray}
The total mass of the shell ($M=A \sigma$) in terms of scalar field
is defined as
\begin{eqnarray}\label{25}
M=4\pi R^2 \sigma= 2\pi R^2[\dot{\psi}^2+2V(\psi)].
\end{eqnarray}
Inserting Eqs.(\ref{24}) and (\ref{25}) in (\ref{14}), we obtain
\begin{eqnarray}\label{26}
\ddot{\psi}+\frac{2\dot{R}}{R}\dot{\psi}+\frac{\partial V}{\partial
\psi}=0.
\end{eqnarray}
This is a well-known Klein-Gordon (KG) equation and its
representation in shell coordinate system is $\Box\psi+\partial
V/\partial \psi=0$.

In order to analyze the dynamical behavior of spherical scalar
shell, we solve KG as well as energy conservation equation
simultaneously for $\psi(\tau)$ and $R(\tau)$. The effective
potential for the Schwarzschild, Kerr and Kerr-Newmann BHs in terms
of scalar field are obtained as
\begin{eqnarray}\label{27}
V_{eff1}(R)&=&-1 + \frac{2 m_{\pm}}{R} +4 \pi^2
R^2\left[\dot{\psi}^2 + 2 V(\psi)\right]^2,
\\\nonumber
V_{eff2}(R)&=&\left\{B \left(B-4A \pi^2
R^2\left[\dot{\psi}^2+2V(\psi)\right]^2\right) \left(A^2F_{\pm}
\left(a^2 + R^2\right)+2 a^2
\right.\right.\\\nonumber&\times&\left.\left. \left(A F_{\pm} +8
m_{\pm} R\right)m_{\pm} R\sin^2\theta \right)\right\}\left\{
\left(A^3 B F_{\pm}\left(a^2 +R^2 \right)+2a^2
\right.\right.\\\nonumber&\times&\left.\left(A B F_{\pm}-32 A
m_{\pm}\pi^2 R^3[\dot{\psi}^2+2V(\psi)]^2 +8 B m
R\right)\right.\\\label{28}&\times&\left.
m_{\pm}AR\sin^2\theta\right\}^{-1},
\\\nonumber
V_{eff3}(R)&=&-\left\{B \left(B-16 \pi^2A R^2\left[\dot{\psi}^2 + 2
V(\psi)\right]^2\right) \left(-A^2F_{\pm} \left(a^2 + R^2\right)
\right.\right.\\\nonumber&+&\left.\left.a^2\left(Q^2-2
m_{\pm}R\right) \left(A F_{\pm}-4Q^2 +8 m_{\pm} R\right)\sin^2\theta
\right)\right\}\left\{\left(A^3 B F_{\pm}
\right.\right.\\\nonumber&\times&\left.\left(a^2 +R^2
\right)-a^2A\left(64 \pi^2A \left(Q^2-2m_{\pm}R\right)
R^2\left[\dot{\psi}^2 + 2 V(\psi)\right]^2
\right.\right.\\\label{29}& +&\left.\left. B\left(A F_{\pm}-4Q^2+8
m_{\pm} R\right)\right)\left(Q^2-2m_{\pm}R\right)\sin^2\theta
\right\}^{-1}.
\end{eqnarray}
Equations (\ref{8}) and (\ref{26}) can be solved numerically for
$m_-=0$ and $m_+=R_0=\sigma_0=k=M=1$. Figure \textbf{4} indicates
the dynamical behavior of shell corresponding to Schwarzschild, Kerr
and Kerr-Newmann BHs. Here, each graph describes the motion of
scalar shell through upper and lower curves indicating the expanding
and collapsing behavior, respectively. For Kerr BH, it is found that
the increase in rotation parameter leads to enhancement of expansion
as well as collapse rate. In case of Kerr-Newmann BH, we analyze
that the presence of charge and rotation parameters decreases the
expansion and collapse rate.
\begin{figure}\centering\label{3}
\epsfig{file=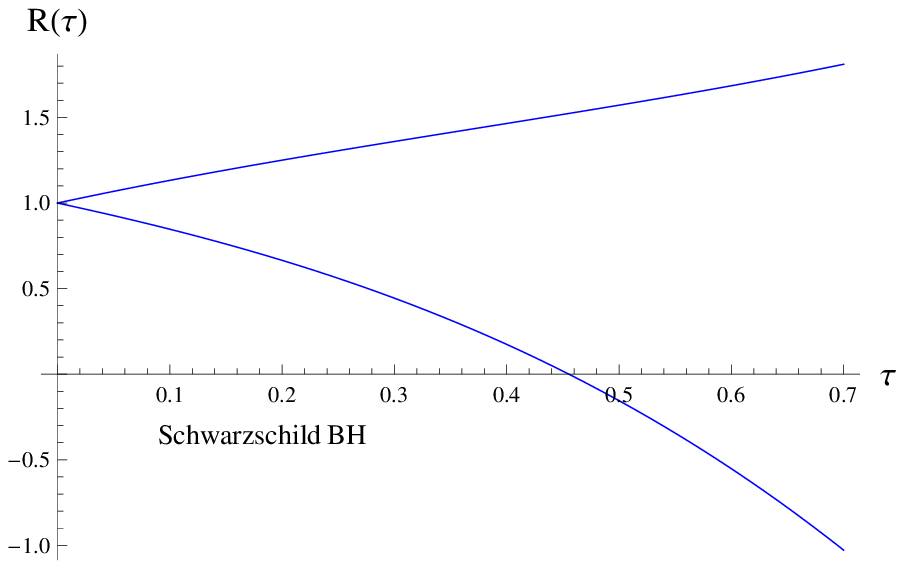,width=.5\linewidth}\epsfig{file=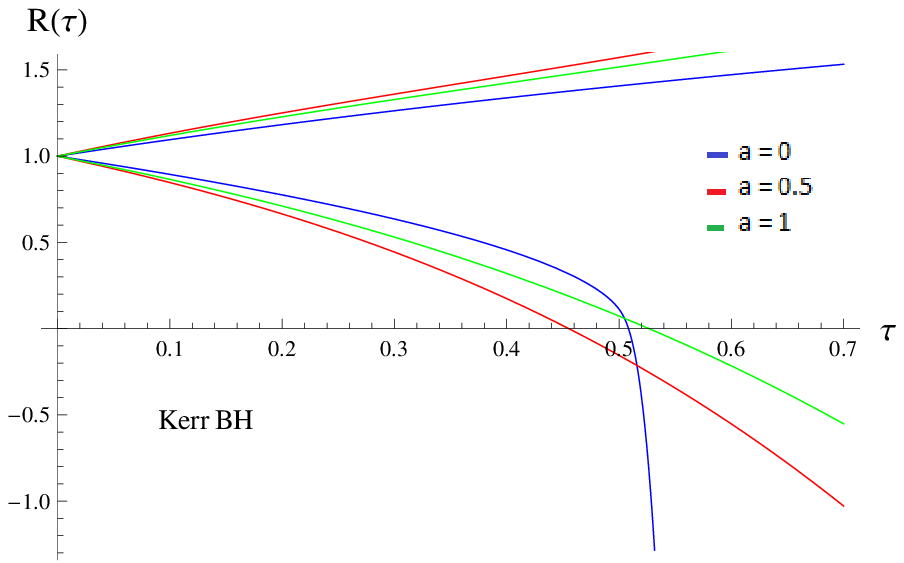,width=.5\linewidth}
\epsfig{file=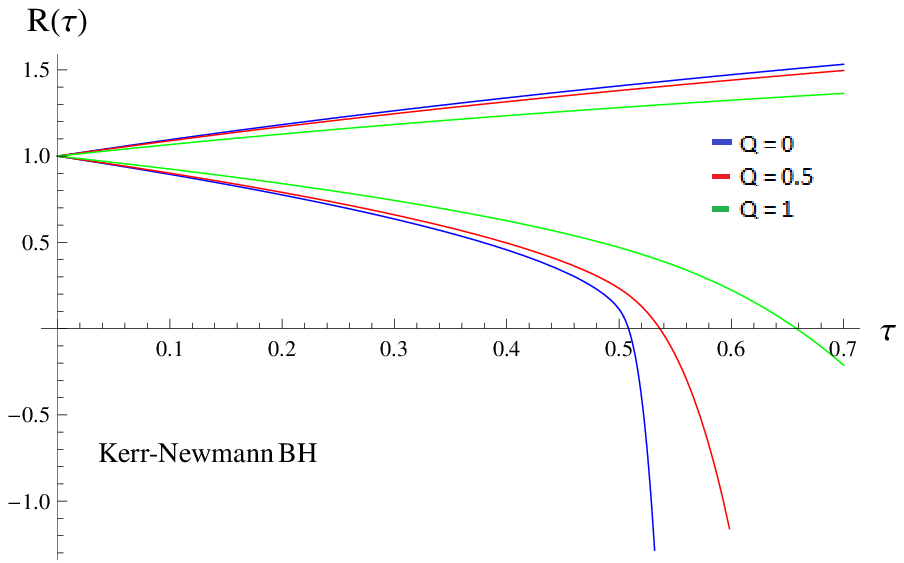,width=.5\linewidth}\epsfig{file=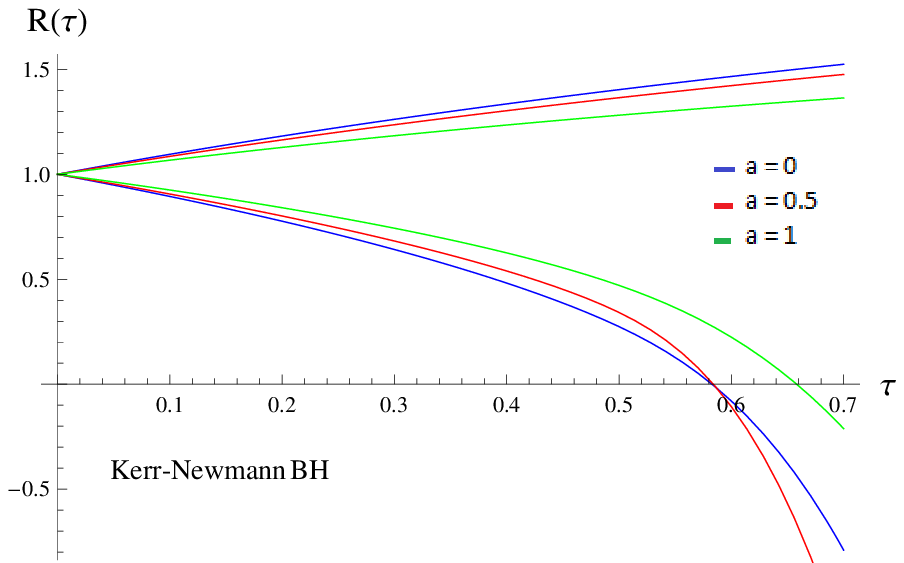,width=.5\linewidth}
\caption{Behavior of shell's radius with scalar field.}
\end{figure}

Now, we study the dynamical behavior of scalar shell for massless as
well as massive scalar field.

\subsubsection{Massless Scalar Shell}

Here we analyze the dynamical behavior of shell in the absence of
scalar potential field $(V(\psi))$, i.e., massless scalar field. In
this case, we do not need different EoS because the vanishing of
$(V(\psi))$ leads to a direct relation between pressure and surface
density ($p=\sigma$). Using the condition $V(\psi)=0$, the KG
equation can be written as
$\ddot{\psi}+\frac{2\dot{R}}{R}\dot{\psi}=0$ and its integration
leads to $\dot{\psi}=\frac{\lambda}{R^2}$, where $\lambda$ is an
integrating constant. The equations of motion for Schwarzschild,
Kerr and Kerr-Newmann BHs become
\begin{eqnarray}\label{30}
\dot{R}^2&=&1 - \frac{2m_{\pm}}{R} - \frac{4 \pi^2 \lambda^4}{R^6},
\\\nonumber \dot{R}^2&=&-\left\{B \left(B-\frac{4A
\pi^2\lambda^4}{R^6}\right) \left(A^2F_{\pm} \left(a^2 +
R^2\right)+2 a^2 m_{\pm} R\sin^2\theta
\right.\right.\\\nonumber&\times&\left.\left. \left(A F_{\pm} +8
m_{\pm} R\right) \right)\right\}\left\{ \left(A^3 B F_{\pm}\left(a^2
+R^2 \right)+2AR a^2 m_{\pm}
\right.\right.\\\label{31}&\times&\left.\left(A B
F_{\pm}-\frac{32\pi^2 A m_{\pm} \lambda^4}{R^5} +8 B m
R\right)\sin^2\theta\right\}^{-1}, \\\nonumber \dot{R}^2&=&\left\{B
\left(B-\frac{16A \pi^2 \lambda^4}{R^6}\right) \left(-A^2F_{\pm}
\left(a^2 + R^2\right)+a^2\left(Q^2-2 m_{\pm}R\right)
\right.\right.\\\nonumber&\times&\left.\left. \left(A F_{\pm}-4Q^2
+8 m_{\pm} R\right)\sin^2\theta \right)\right\}\left\{ \left(A^3 B
F_{\pm}\left(a^2 +R^2 \right)-a^2A
\right.\right.\\\nonumber&\times&\left.\left(\frac{64\pi^2  A
\left(Q^2-2m_{\pm}R\right) \lambda^4}{R^6} + B\left(A F_{\pm}-4Q^2+8
m_{\pm} R\right)\right)\right.\\\label{32}&
\times&\left.\left(Q^2-2m_{\pm}R\right)\sin^2\theta \right\}^{-1}.
\end{eqnarray}
The corresponding effective potential are
\begin{eqnarray}\label{33}
V_{eff1}(R)&=&-1 + \frac{2m_{\pm}}{R} + \frac{4 \pi^2
\lambda^4}{R^6}, \\\nonumber V_{eff2}(R)&=&\left\{B \left(B-\frac{4A
\pi^2\lambda^4}{R^6}\right) \left(A^2F_{\pm} \left(a^2 +
R^2\right)+2 a^2 m_{\pm} R\sin^2\theta
\right.\right.\\\nonumber&\times&\left.\left. \left(A F_{\pm} +8
m_{\pm} R\right) \right)\right\}\left\{ \left(A^3 B F_{\pm}\left(a^2
+R^2 \right)+2AR a^2 m_{\pm}
\right.\right.\\\label{34}&\times&\left.\left(A B
F_{\pm}-\frac{32\pi^2 A m_{\pm} \lambda^4}{R^5} +8 B m
R\right)\sin^2\theta\right\}^{-1},
\\\nonumber
V_{eff3}(R)&=&-\left\{B \left(B-\frac{16A \pi^2
\lambda^4}{R^6}\right) \left(-A^2F_{\pm} \left(a^2 +
R^2\right)+\left(Q^2-2 m_{\pm}R\right)
\right.\right.\\\nonumber&\times&\left.\left. \left(A F_{\pm}-4Q^2
+8 m_{\pm} R\right)a^2\sin^2\theta \right)\right\}\left\{ \left(A^3
B F_{\pm}\left(a^2 +R^2 \right)-a^2A
\right.\right.\\\nonumber&\times&\left.\left(\frac{64\pi^2  A
\left(Q^2-2m_{\pm}R\right) \lambda^4}{R^6} + B\left(A F_{\pm}-4Q^2+8
m_{\pm} R\right)\right)\right.\\\label{35}&
\times&\left.\left(Q^2-2m_{\pm}R\right)\sin^2\theta \right\}^{-1}.
\end{eqnarray}
\begin{figure}\centering
\epsfig{file=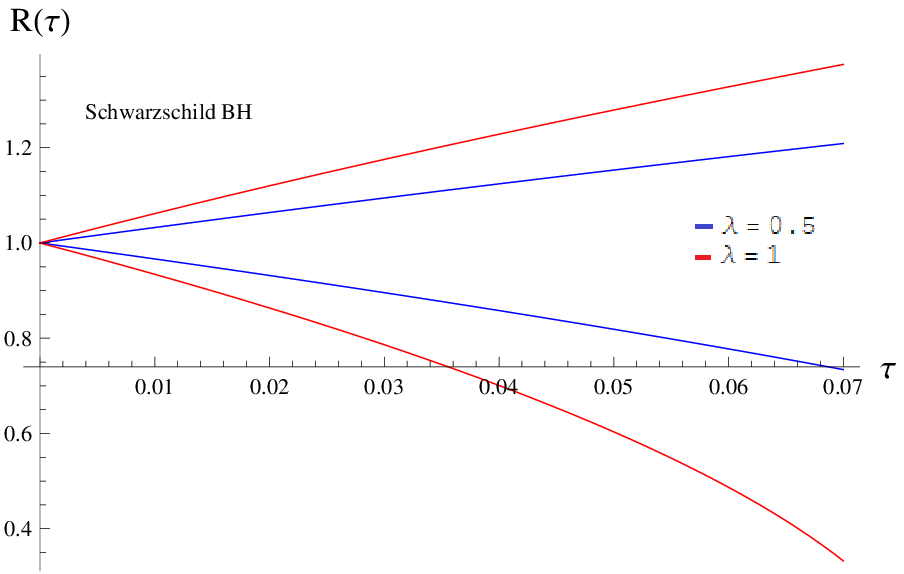,width=.5\linewidth}\epsfig{file=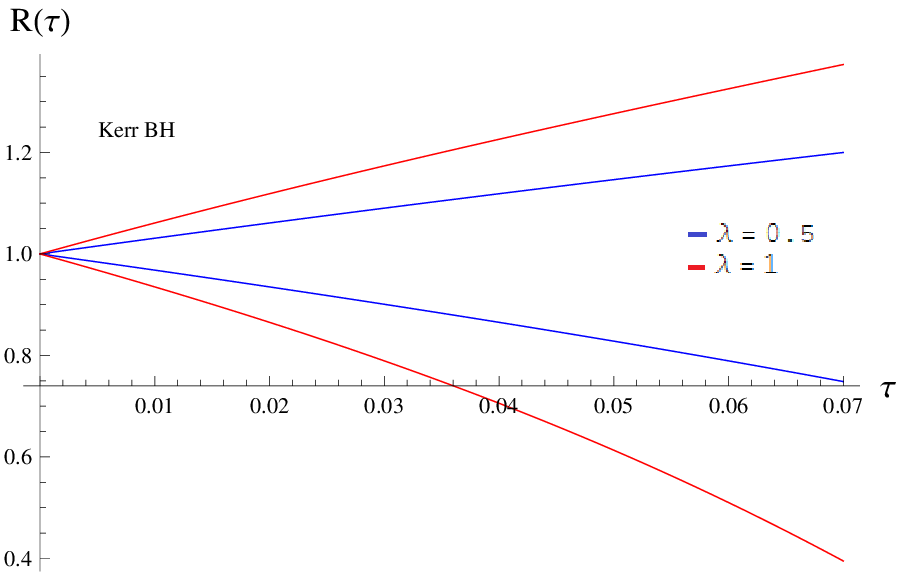,width=.5\linewidth}
\epsfig{file=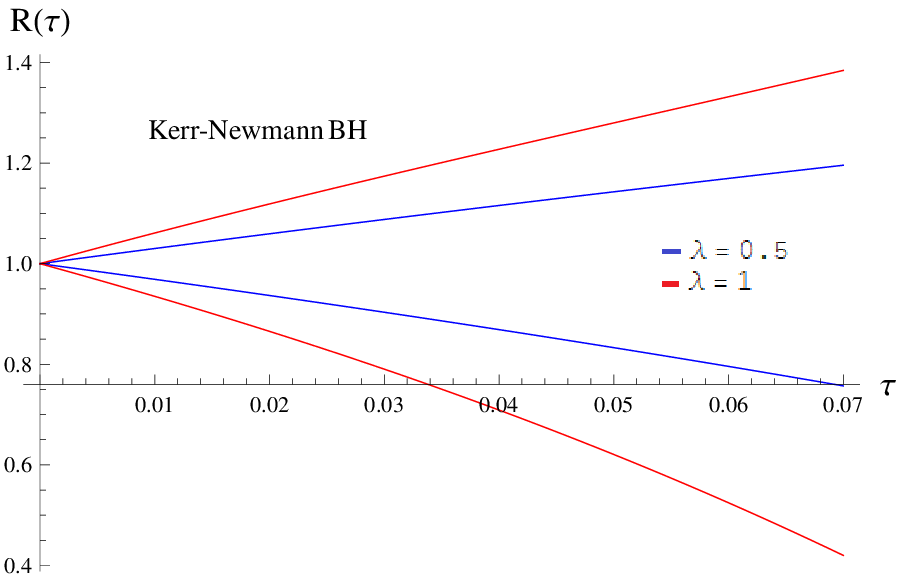,width=.5\linewidth}\epsfig{file=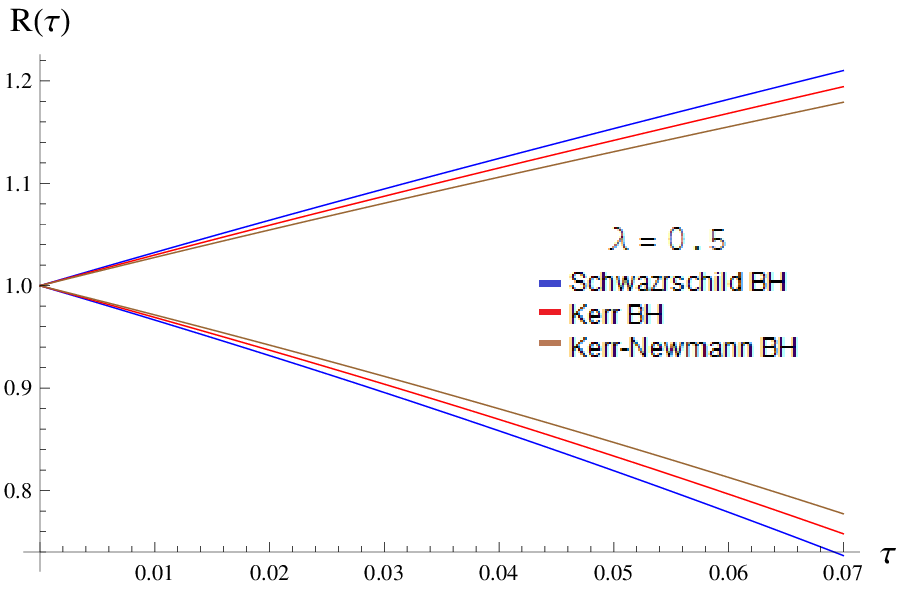,width=.5\linewidth}
\caption{Behavior of shell's radius with massless scalar field.}
\end{figure}

The radius of shell and effective potential for massless scalar
field are plotted in Figures \textbf{5-7}. Figure \textbf{5} shows
increasing and decreasing behavior of radius which leads to
expansion and collapse, respectively. It is also observed that the
rate of expansion as well as collapse increases by increasing
$\lambda$. The right plot of lower panel shows that the contribution
of charge and rotation parameters decreases the collapse as well as
expansion rate. Figures \textbf{6} and \textbf{7} represent the
dynamics of massless scalar shell through effective potential. These
plots describe the expansion ($V_{eff}>0$), collapse ($V_{eff}<0$)
and saddle points ($V_{eff}=0$) of massless scalar shell. Figures
\textbf{6} and \textbf{7} show that the scalar shell has the same
behavior of expansion, collapse as well as saddle points. It is
shown that the dynamical behavior of shell decreases by enhancing
charge and rotation parameters.
\begin{figure}\centering
\epsfig{file=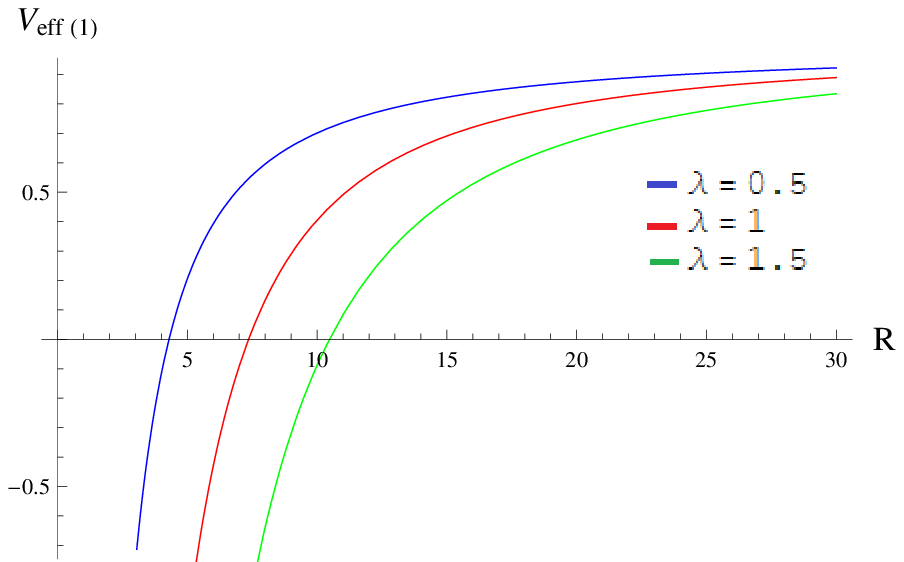,width=.5\linewidth}\epsfig{file=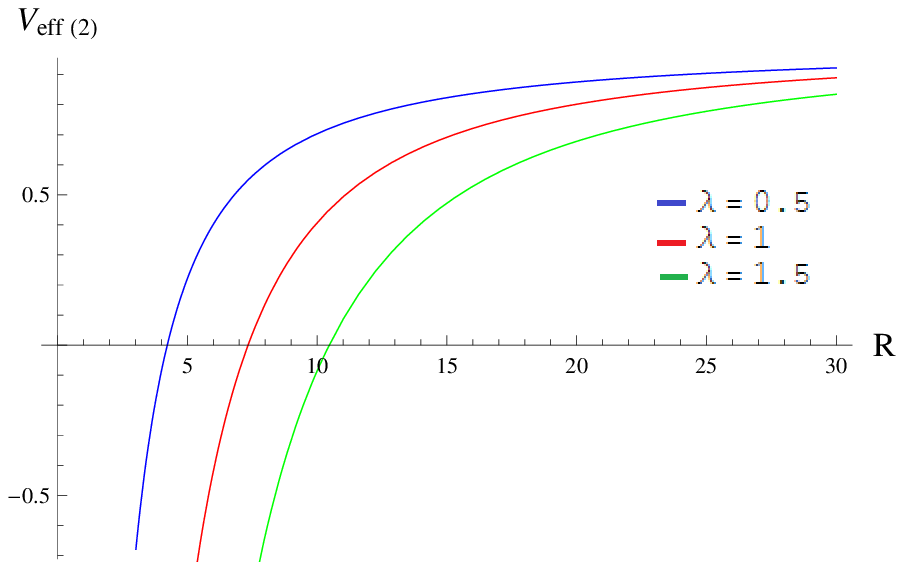,width=.5\linewidth}
\epsfig{file=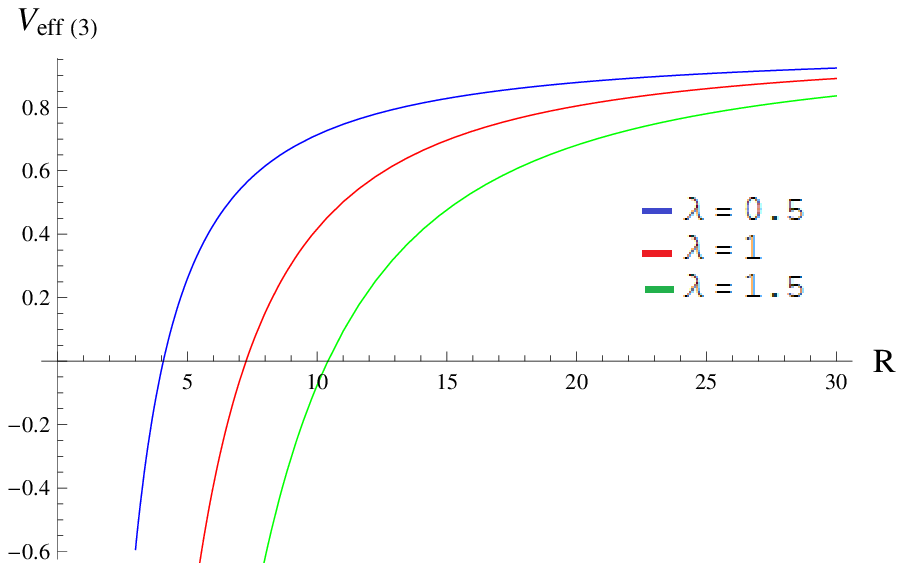,width=.5\linewidth} \caption{Plots of
$V_{eff}$ versus $R$ for the interior region.}
\end{figure}
\begin{figure}\centering
\epsfig{file=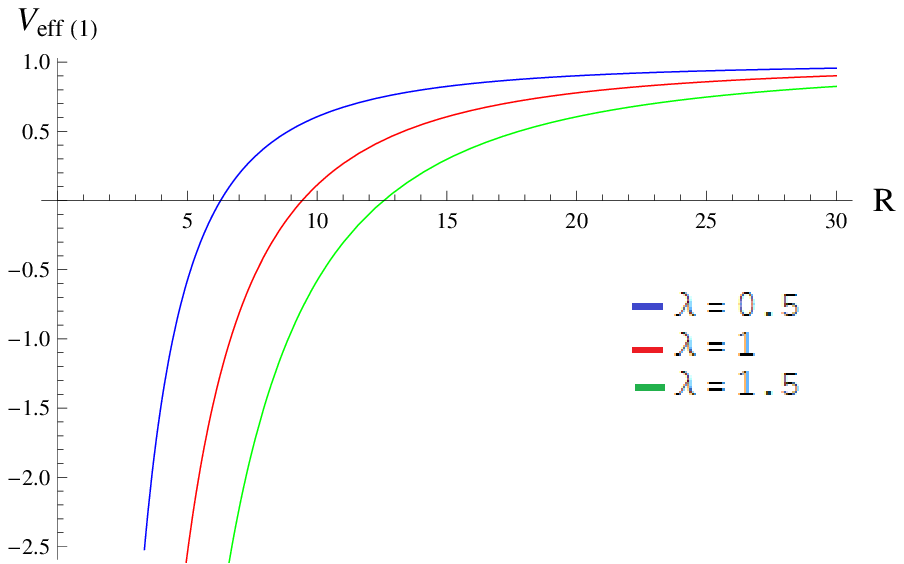,width=.5\linewidth}\epsfig{file=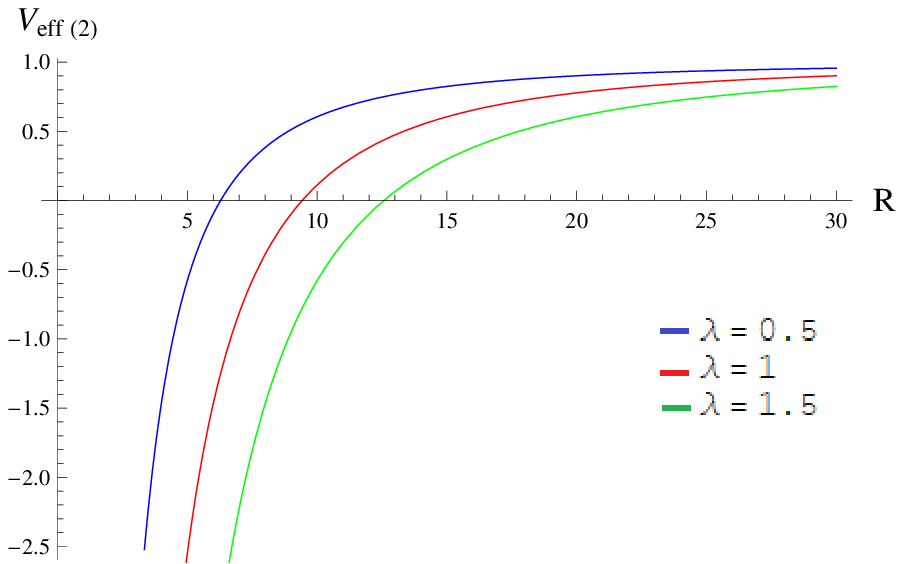,width=.5\linewidth}
\epsfig{file=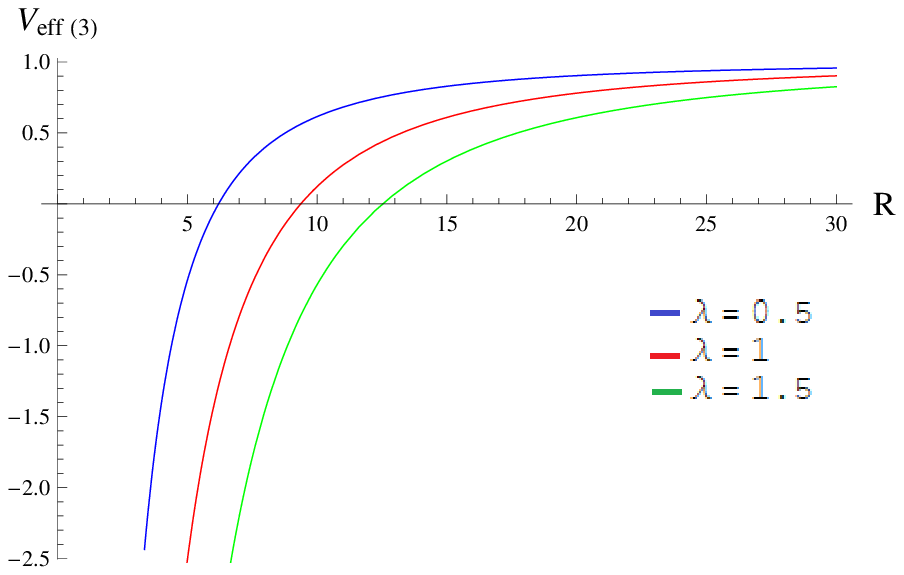,width=.5\linewidth} \caption{Plots of
$V_{eff}$ versus $R$ for the exterior region.}
\end{figure}

\subsubsection{Massive Scalar Shell}

In this case, we discuss the dynamics of scalar shell through
massive scalar field, i.e., $V(\psi)=M^2\psi^2$. From Eq.(\ref{24}),
we find the potential function and massive scalar field given as
\begin{eqnarray}\label{36}
\dot{\psi}^2=\sigma+p, \quad 2V(\psi)=p-\sigma .
\end{eqnarray}
We consider surface pressure as explicit function of $R$, i.e.,
$p=p_{0}e^{-\gamma R}$, where $\gamma$ and $p_{0}$ are constants.
Using Eq.(\ref{14}) alongwith the value of $p$, we find
\begin{eqnarray}\label{37}
\sigma=\frac{\omega}{R^2}+\frac{2\left(1+\gamma
R\right)p_{0}e^{-\gamma R}}{\gamma^2 R^2},
\end{eqnarray}
where $\omega$ is an integration constant. Using the values of
energy density and surface pressure in Eq.(\ref{36}), we obtain
\begin{eqnarray}\label{38}
V(\psi)&=&\frac{\omega}{2R^2}-\frac{p_{0}e^{-\gamma
R}}{2}\left(1-\frac{2(1+\gamma R)}{\gamma^2 R^2}\right),
\\\label{39}
\dot{\psi}^2&=&\frac{\omega}{R^2}-p_{0}e^{-\gamma
R}\left(1+\frac{2\left(1+\gamma R\right)}{\gamma^2 R^2}\right),
\end{eqnarray}
which satisfy the KG equation. Using Eqs.(\ref{37})-(\ref{39}) in
(\ref{27})-(\ref{29}), it follows that
\begin{eqnarray}\label{40}
V_{eff1}(R)&=&-1+ \frac{M^2}{R^2} +\frac{2m_{\pm}}{R}, \\\nonumber
V_{eff2}(R)&=&\left\{B \left(B-\frac{A M^2}{R^2}\right)
\left(A^2F_{\pm} \left(a^2 + R^2\right)+2 a^2 m_{\pm} R\sin^2\theta
\right.\right.\\\nonumber&\times&\left.\left. \left(A F_{\pm} +8
m_{\pm} R\right) \right)\right\}\left\{ \left(A^3 B F_{\pm}\left(a^2
+R^2 \right)+2AR a^2 m_{\pm}
\right.\right.\\\label{41}&\times&\left.\left(A B F_{\pm}-\frac{8 A
m_{\pm} M^2}{R} +8 B m R\right)\sin^2\theta\right\}^{-1},
\\\nonumber V_{eff3}(R)&=&-\left\{B \left(B-\frac{4A M^2}{R^2}\right)
\left(-A^2F_{\pm} \left(a^2 + R^2\right)+a^2\left(Q^2-2
m_{\pm}R\right) \right.\right.\\\nonumber&\times&\left.\left.
\left(A F_{\pm}-4Q^2 +8 m_{\pm} R\right)\sin^2\theta
\right)\right\}\left\{ \left(A^3 B F_{\pm}\left(a^2 +R^2
\right)-a^2A \right.\right.\\\nonumber&\times&\left.\left(\frac{16 A
\left(Q^2-2m_{\pm}R\right) M^2}{R^2} + B\left(A F_{\pm}-4Q^2+8
m_{\pm} R\right)\right)\right.\\\label{42}&
\times&\left.\left(Q^2-2m_{\pm}R\right)\sin^2\theta \right\}^{-1}.
\end{eqnarray}
Also, the mass of shell can be expressed as
\begin{eqnarray}\label{43}
M=4\pi R^2\sigma=4\pi\omega+\frac{8\pi p_{0}e^{-\gamma
R}}{\gamma^2}\left(1+\gamma R\right).
\end{eqnarray}
\begin{figure}\centering
\epsfig{file=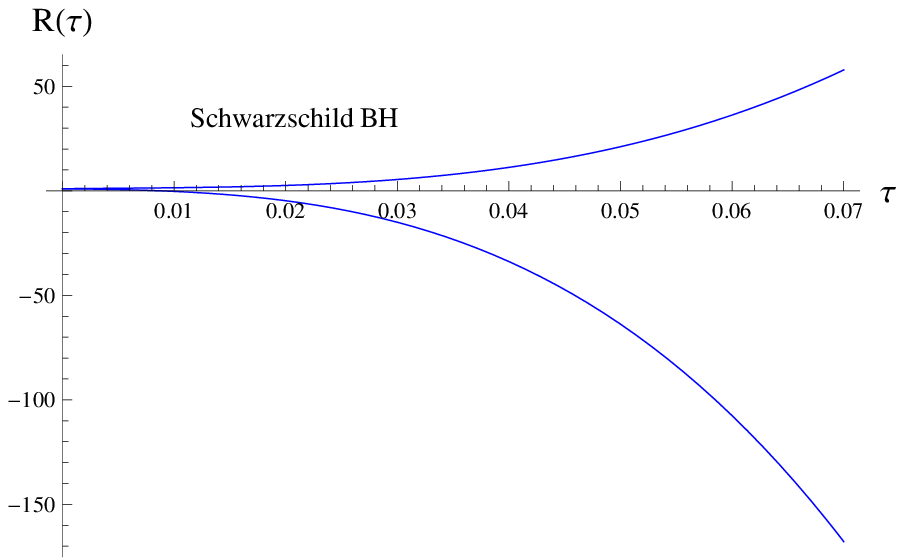,width=.5\linewidth}\epsfig{file=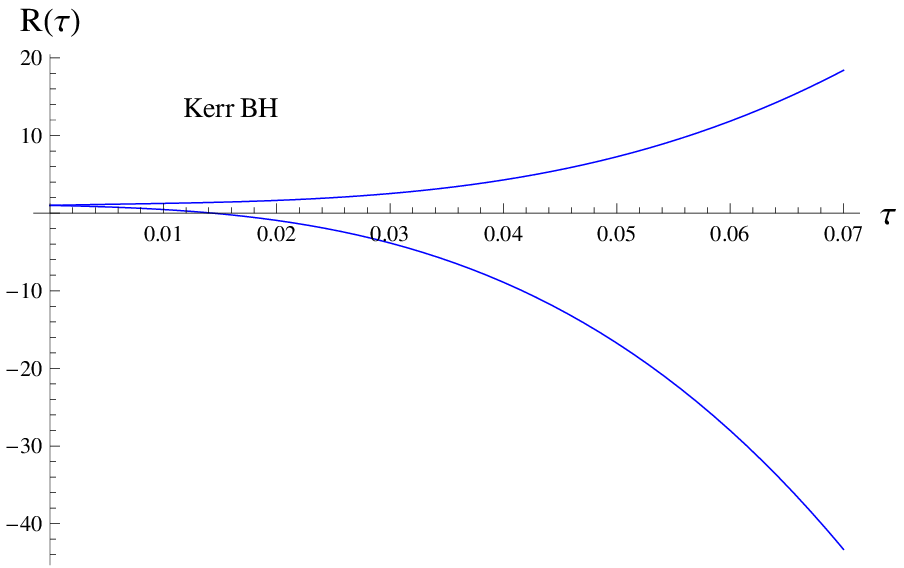,width=.5\linewidth}
\epsfig{file=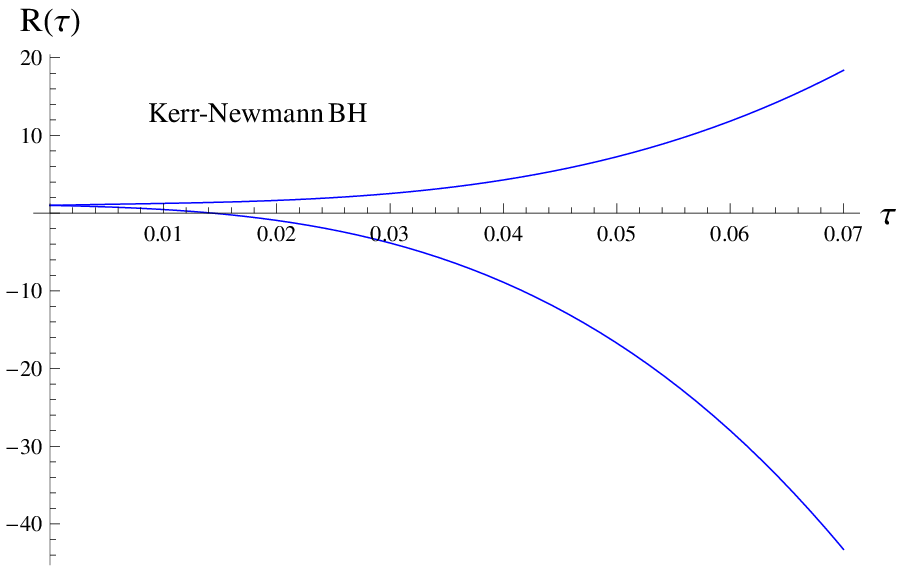,width=.5\linewidth} \caption{Behavior of
shell's radius with massive scalar shell.}
\epsfig{file=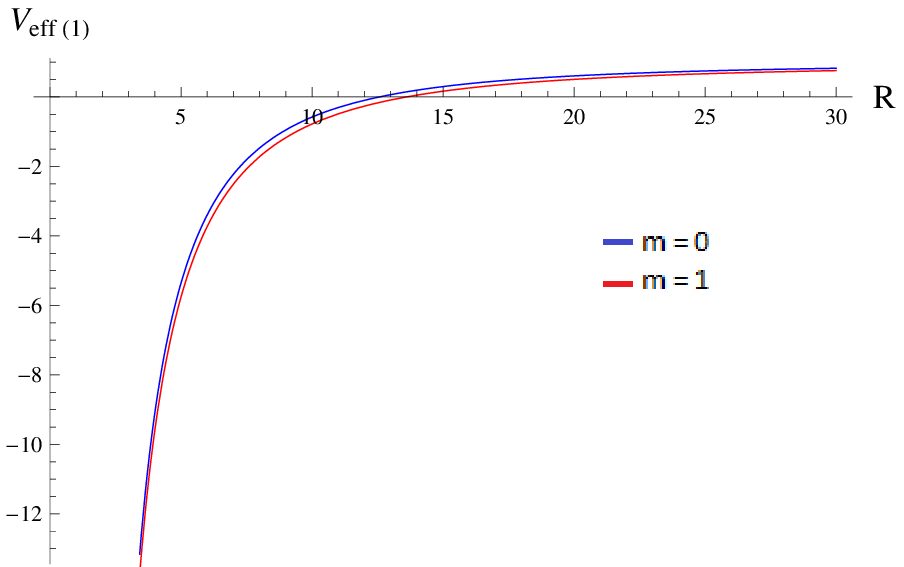,width=.5\linewidth}\epsfig{file=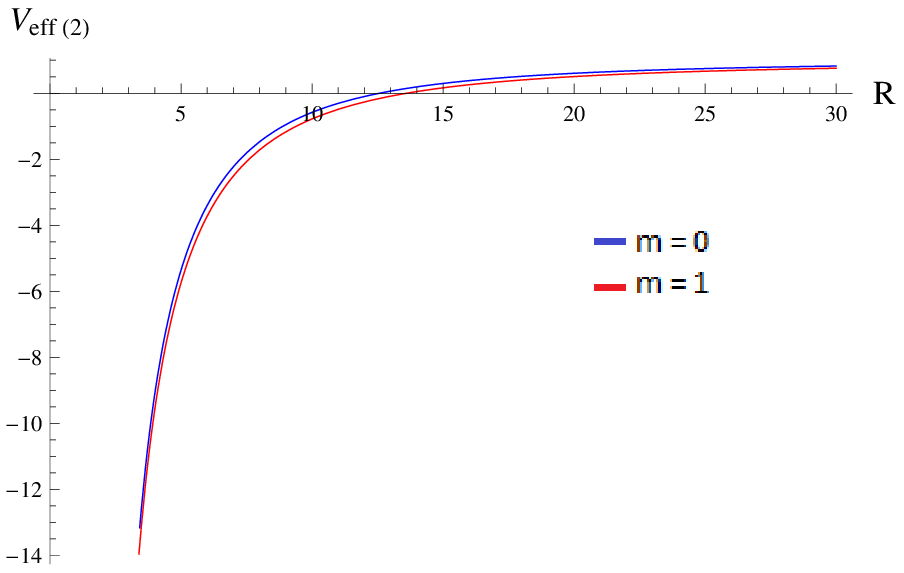,width=.5\linewidth}
\epsfig{file=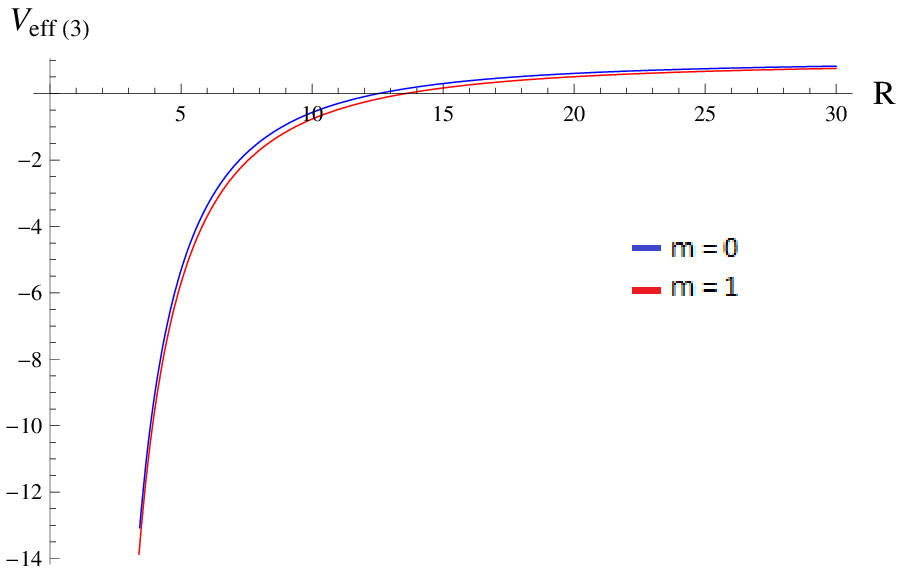,width=.5\linewidth} \caption{Plots of $V_{eff}$
versus $R$ for massive case when $\omega=1$.}
\end{figure}
\begin{figure}\centering
\epsfig{file=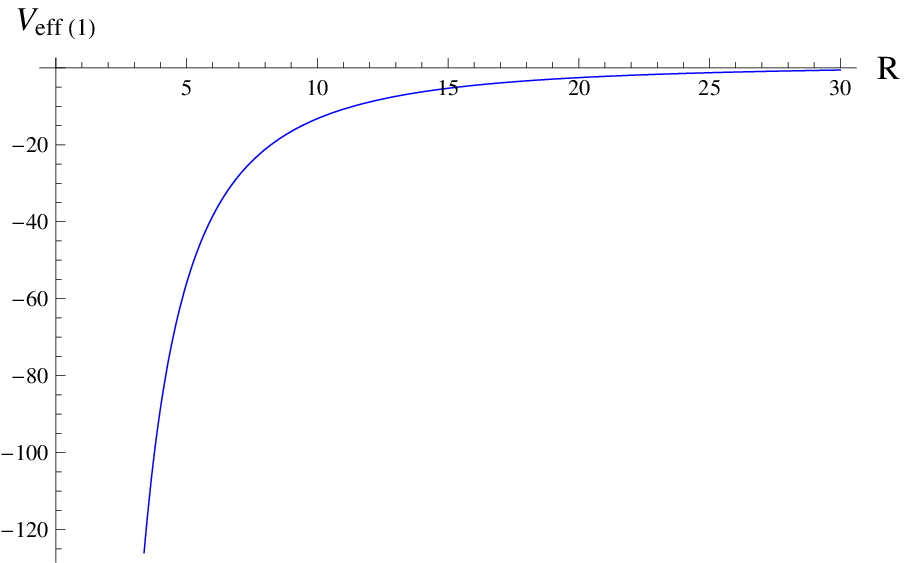,width=.5\linewidth}\epsfig{file=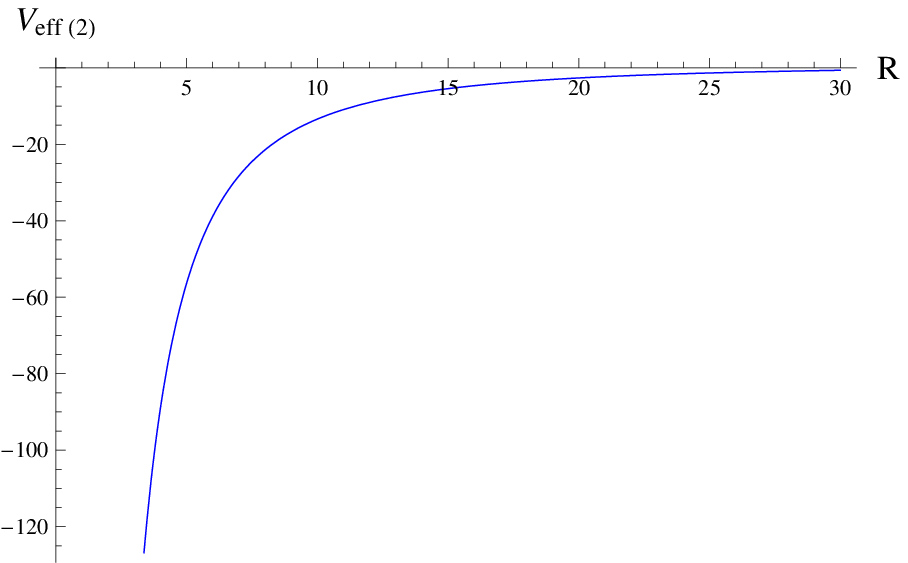,width=.5\linewidth}
\epsfig{file=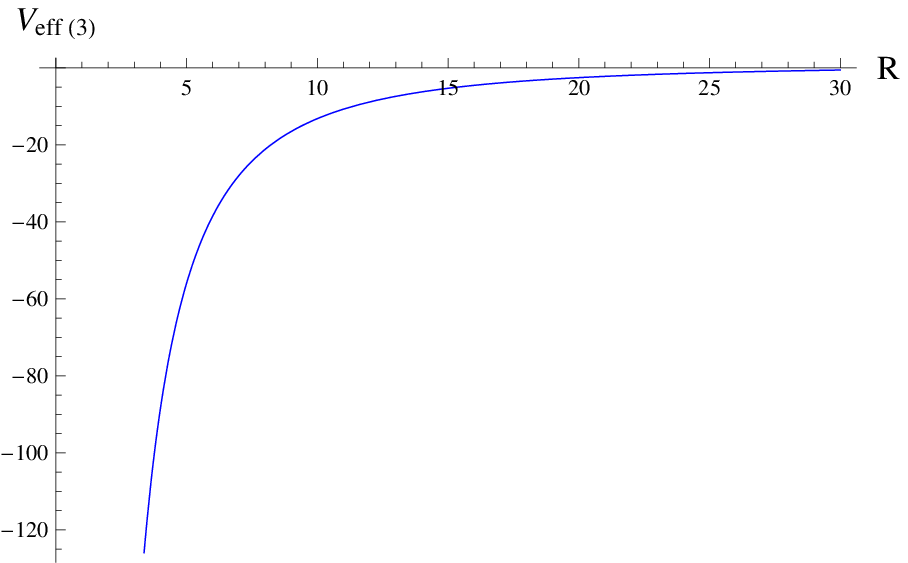,width=.5\linewidth} \caption{Plots of $V_{eff}$
versus $R$ for massive case when $\omega=3$ .}
\end{figure}

Figures \textbf{8-10} indicate the behavior of shell's radius as
well as effective potential for massive scalar field for $m_-=0$ and
$m_+=R_0=\sigma_0=\omega=\gamma=Q=a=1$. In Figure \textbf{8}, the
upper curve represents the expansion of scalar shell while the lower
curve shows collapse. Figure \textbf{9} shows that the effective
potential for massive scalar field leads to expansion, collapse as
well as saddle points for $\omega\in(-2,2)$. It is also found that
$-2\geq\omega\geq2$ leads the shell to collapse (negative effective
potential) only (Figure \textbf{10}). For massive scalar field, we
find that the effective potential exhibits similar behavior
(expansion, collapse and saddle point) for all values of free
parameters.

\section{Final Remarks}

In this paper, we have studied the dynamics of scalar shell for a
class of BHs using Israel thin-shell formalism. For this purpose, we
have formulated the equations of motion which leads to the behavior
of shell's velocity. We have then investigated the dynamics of
scalar shell for massless and massive scalar fields. The results can
be summarized as follows.
\begin{itemize}
\item The motion of scalar shell shows expanding as well as
collapsing behavior of the shell (Figures \textbf{1-4}). It is found
that shell's velocity has similar behavior for all considered BHs.
For the Kerr-Newmann BH, it is analyzed that the velocity of shell
decreases by enhancing charge parameter (Figure \textbf{3}). We have
obtained that the rate of expansion and collapse for Schwarzschild
BH is greater than the Kerr and Kerr-Newmann BHs (Figure
\textbf{4}).
\item For massless scalar field, it is
observed that the presence of charge and rotation parameters affect
the dynamical behavior of shell's radius. For the Kerr and
Kerr-Newmann BHs, expansion as well as collapse rate is small as
compared to the Schwarzschild BH (Figure \textbf{5}). The behavior
of effective potential indicates the expansion, collapse as well as
the stable points (Figures \textbf{6-7}). For both interior and
exterior spacetimes, expansion and collapse rates are decreased as
$\lambda$ increases (Figures \textbf{6} and \textbf{7}).
\item For massive scalar field, it is shown that
the radius of shell either expands or undergoes collapse. We have
found that Schwarzschild BH exhibits more expansion and collapse as
compared to Kerr and Kerr-Newmann BHs. This describes that the
contribution of charge and rotation parameters decreases these
phenomena (Figure \textbf{8}). It is found that $V_{eff}$ shows
expansion as well as collapse of thin-shell for $\omega\in(-2,2)$
while it indicates only collapse for other values of $\omega$
(Figures \textbf{9} and \textbf{10}). It is observed that the
behavior of effective potential for three BHs overlaps for all
values of charge as well as rotation parameter.
\end{itemize}
We conclude that the dynamical evolution of scalar shell can be
expressed through continuous expansion, collapse and stable
configuration.

\end{document}